# PARSIR: a Package for Effective Parallel Discrete Event Simulation on Multi-processor Machines

Francesco Quaglia
DICII - University of Rome Tor Vergata
francesco.quaglia@uniroma2.it

*Abstract*—In this article we present PARSIR (PARallel SImulation Runner), a package that enables the effective exploitation of shared-memory multi-processor machines for running discrete event simulation models. PARSIR is a compile/run-time environment for discrete event simulation models developed with the C programming language. The architecture of PARSIR has been designed in order to keep low the amount of CPU-cycles required for running models. This is achieved via the combination of a set of techniques like: 1) causally consistent batch-processing of simulation events at an individual simulation object for caching effectiveness; 2) high likelihood of disjoint access parallelism; 3) the favoring of memory accesses on local NUMA (Non-Uniform-Memory-Access) nodes in the architecture, while still enabling well balanced workload distribution via work-stealing from remote nodes; 4) the use of RMW (Read-Modify-Write) machine instructions for fast access to simulation engine data required by the worker threads for managing the concurrent simulation objects and distributing the workload. Furthermore, any architectural solution embedded in the PARSIR engine is fully transparent to the application level code implementing the simulation model. We also provide experimental results showing the effectiveness of PARSIR when running the reference PHOLD benchmark on a NUMA shared-memory multi-processor machine equipped with 40 CPUs.

## I. INTRODUCTION

The class of techniques that allow a Discrete Event Simulation (DES) model to be run in a parallel/distributed environment is known as Parallel DES (PDES) [1]. It is based on partitioning the simulation model into multiple *simulation objects*, which can be executed concurrently on the underlying computing system. At the same time, the simulation objects are related to each other thanks to the cross-scheduling of simulation events. This enables modeling that something occurs in a part of the model depending on what occurred in some other part of it.

A key concept for PDES is "event causality" which states that any individual simulation object needs to process its incoming events in non-decreasing order of their logical time occurrence. This is also referred to as *event timestamp*. As for this concept, a key factor exploited in the literature is the notion of *lookahead* of the simulation model. It measures the lower-bound distance between the timestamp of an event $A$ and the timestamp of any other event $B$ possibly generated while processing $A$ at some simulation object. Hence, it can be exploited for determining in what part of the simulation time-line no additional event can be further generated while processing events that have already been scheduled to occur. This enables determining what events can be processed at any simulation object without violations of the causality property. Given its central importance in PDES, several studies have addressed how to determine improved values of the simulation model lookahead (see, e.g., [2]–[4]).

However, as also discussed in [5], the design and development of PDES engines is far from being a trivial task since the potential dependencies across the simulation objects, and the dynamism of event production along the model execution, make this type of parallel programming context extremely complex. Furthermore, a modern PDES package should consider peculiarities in the underlying hardware platform, hence embedding specific optimizations for the management of performance critical aspects at the hardware level, and their relations with the software structure of the simulation engine.

In this article, we present PARSIR (PARallel SImulation Runner), an open source package available to the community [6] which implements advanced technical solutions for enabling the effective exploitation of shared-memory multi-processor machines for PDES. It has been designed and realized for Posix operating systems (e.g., Linux) but it could be easily ported to other types of operating systems (e.g., Windows systems). Also, it offers a compile/run-time environment for simulation models developed using the C programming language.

PARSIR is based on a combination of a set of techniques that are oriented to the reduction of the number of CPU-cycles required for running the overlying DES model. In particular, it targets:

1) the reduction of the incidence of cache misses, which is achieved thanks to batch-processing of the events at the different simulation objects, while still ensuring causal consistency;
2) the reduction of the number of CPU-cycles spent for waiting other threads while accessing shared data at the level of the simulation engine, which is achieved thanks to data-structure organization leading to high likelihood of disjoint access parallelism;
3) the controlled placement of simulation objects on different NUMA nodes, combined with the prioritized access of threads to objects on local NUMA nodes;
4) the redirection of thread accesses to remote NUMA nodes, via workload stealing, if local accesses are no longer executable without risks of causality violations;

5) the reliance on atomic RMW (Read-Modify-Write) machine instructions—and the avoidance of critical sections—for the dynamic workload distribution within a same NUMA node, and the stealing of workload from remote nodes.

PARSIR is based on compile and run-time techniques which are typical of system-level layers. Also, it fully masks all of its parallelism and hardware-asymmetry oriented solutions to the overlying application code. It has therefore an extremely wide usability for the development of any kind of DES model to be transparently run in parallel on a shared-memory multi-processor machine, with no need for the programmer to face aspects related to the underlying platform. Actually, the interaction between the engine-level software of PARSIR and the overlying application software is based on the following API: a `ScheduleNewEvent(...)` service, callable by the application for injecting (i.e., scheduling) a new event to be processed by whichever simulation object in the model, and a `ProcessEvent(...)` callback offered by the application in order to enable the simulation engine to pass control to the simulation-model specific code for processing an event that was previously scheduled to occur. These two services are classical in the literature of (P)DES. Therefore, PARSIR simply aligns itself to this widely employed best practice.

Finally, it is important to note that, while presenting PARSIR, we do not intend to provide innovative methods for PDES. In fact, PARSIR is essentially based on a traditional synchronous PDES algorithm with constant global lookahead [1]. Rather, we illustrate the careful realization of a framework that combines several technical aspects in a smart manner, supporting hardware-effective execution of the overlying simulation model, in particular in terms of avoidance/reduction of stale CPU-cycles and hardware level traffic that may obstacle performance. Also, experimental results providing indications on the opportunity of using PARSIR are reported.

The remainder of this article is structured as follows. Section II provides the description of the main architectural choices in the PARSIR package. Its relation with the literature is discussed in Section III. Results of runs with PARSIR, showing to the reader its main capabilities, are provided in Section IV. Conclusions are discussed in Section V.

## II. THE PARSIR PACKAGE

### A. Epochs and Workload Distribution

As we already noted, PARSIR is a shared-memory multi-processor oriented implementation of a synchronous PDES algorithm with constant global lookahead [1]. In particular, PARSIR uses the simulation-model lookahead, which we denote as $L$, to devise the concept of epoch of the PDES application execution. The $i$-th epoch is the simulation time interval $[i \times L, (i+1) \times L)$ that contains all the events of the application such that, for each event $e$, its timestamp $T(e)$ satisfies the following inequality

$$i \times L \leq T(e) < (i+1) \times L \qquad (1)$$

with $i \geq 0$. We indicate with $E_i$ the $i$-th epoch of the model execution. In PARSIR, no thread can start processing events that belong to epoch $E_{i+1}$ (or a successive one) unless all the events of the model that stand into epoch $E_i$ are already processed.

As for cache efficiency, PARSIR exploits the fact that, when some thread processes any event $e \in E_i$, no new event $e'$ will be injected in the simulation within the same epoch $E_i$. The immediate consequence is that, when a thread starts processing an event destined to whichever simulation object $o$ during epoch $E_i$, it can process all the events in $E_i$ destined to $o$ in a batch, ordered according to their timestamps, with no need to switch its activity to other objects (even if they have pending events with lower timestamps)—in fact, the switch would simply reduce the level of locality of the thread operations, contrasting caching effectiveness. Event causality is anyhow ensured for object $o$ under such batch execution, even if it gives rise to non-fully (cross-object) ordered processing of the events into the epoch $E_i$ on the basis of their timestamps.

Overall, batch execution enables high cache exploitation since, once the data structures embedding the state of the simulation object $o$ are accessed for processing some event, they are likely found in cache while processing any subsequent event of the same object along the batch. In other words, the object becomes hot when starting processing its events in the epoch $E_i$, and remains hot up to the completion of all its events belonging to that epoch.

One point the reader can note is that the locality for processing the events at an individual simulation object also depends on the way these events are retrieved from the event pool. This aspect, in particular the structure of event queues in PARSIR, will be dealt with in Section II-B.

Concerning other two core points—namely the improvement of cache-miss management, in particular in NUMA machines, and the reduction of the waiting-time spent by any thread in the synchronization barrier that takes place at the end of the processing of each epoch—PARSIR distributes the workload of events according to the below described scheme. As a preliminary note, PARSIR pins each of its worker threads to a specific CPU, belonging to a given NUMA node, and avoids executions with more threads than CPUs. This is a classical approach when using multithreading in the context of PDES applications, which can have very fine-grain events, such that the context switch across different threads on a same CPU for supporting the execution might give rise to excessive housekeeping costs.

The simulation objects are distributed by PARSIR across the NUMA nodes where the worker threads are running. This is done automatically by PARSIS at startup time, by simply packing the identifiers of the simulation objects in knapsacks, each one targeting a specific NUMA node. We indicate with `min[i]` and `max[i]` the minimum and maximum identifiers of objects that are assigned to the NUMA node $i$. PARSIR exploits these values in combination with a per NUMA node counter `c[i]` initialized to 0 at the start of each epoch, and makes any thread acquire workload according to the following

algorithm

```
for(i = LOCAL_NUMA_NODE, j = 0 ; j < NUMA_NODES;){
    target = __sync_fetch_and_add(&c[i%NUMA_NODES],1);
    if(target + min[i%NUMA_NODES] <= max[i%NUMA_NODES]){
        //we found a valid object identifier
        //for the current NUMA node
        ProcessCurrentEpochEvents(target+min[i%NUMA_NODES]);
    }
    else{//move to the next NUMA node
        i++;
        j++;
    }
}
```

In the above algorithm, the `__sync_fetch_and_add()` intrinsic offered by `gcc`—which is internally built using a RMW atomic machine instruction, like the `XADD` instruction prefixed by `LOCK` on x86 processors—is used to uniquely assign an object identifier possibly valid on a given NUMA node to the caller thread. It is interesting to note that the assignment of an object to a thread is dynamic, does not require any critical section for updating shared data structures (since the fetch-and-add technique simply builds on a unique machine instruction executed atomically over the memory hierarchy) and does not eliminate more than a single object identifier to process at each cycle (thus leaving any other object to process to other threads that can become free for taking it).

At the same time, each thread initially acquires the object identifiers that can be valid for the local NUMA node, just by exploiting the `LOCAL_NUMA_NODE` macro, which provides the identifier of the local NUMA node starting from the identifier of the CPU where the thread is pinned. For each valid object identifier, the thread processes in a batch all its events belonging to the current epoch. As soon as no valid object identifiers are still available for the local NUMA node, the thread starts picking identifiers from remote NUMA nodes.

Through this workload distribution algorithm we have higher priority of accesses to objects on the local NUMA node, while being able to steal workload from remote nodes when threads have no longer work to perform in the current epoch from the local NUMA node. This not only favors the memory access latency when cache misses take place (thanks to prioritization of accesses to the local NUMA node) but also enables reducing the cross thread wait phases at the end of the for-cycle before they can move to process the next epoch—just thanks to the cross-thread help provided for processing the workload from remote NUMA nodes. This orients the solution to work conserving the CPUs (namely, high likelihood of actual exploitation of the available CPU cycles for processing any epoch). Clearly, the algorithm still works correctly over machines having a single memory node (i.e., `NUMA_NODES` equal to 1), which might represent a more restricted/limited architectural setup for the case of shared-memory platforms.

As a final note, in the algorithm we only rely on object identifiers, assuming that each object is a set of data structures guaranteed to be placed in memory according to some NUMA rule. How PARSIR manages the actual placement of the different objects on the different NUMA nodes, thus providing the core background for the aforementioned workload distribution algorithm, is discussed in Section II-C.

## B. Event Queues

A central data structure for the management of the epoch-based progression of the simulation is the pool of pending events. It keeps all the event buffers that have information to be provided to the `ProcessEvent(...)` callback used to pass control to the simulation objects. These event buffers are in turn inserted in the event pool when some simulation object has control along a thread and calls the `ScheduleNewEvent(...)` service offered by the simulation platform. In PARSIR, the event pool is a multi-queue, with any individual queue associated with a specific object identifier. In turn, the $i$-th queue in the multi-queue data structure is a calendar with $N$ different buckets, each corresponding to an epoch $E_i$. The queue covers therefore a simulation time interval of size $N \times L$, being the lookahead $L$ the breadth of an individual epoch.

With this organization, once a thread picks the identifier of an object according to the workload distribution algorithm presented in Section II-A, it can extract from the associated calendar (i.e., from the bucket associated with the current epoch) all the events belonging to the current epoch and destined to this object without any need for traversing pointer-linked event-buffers destined to other objects. This again favors locality in the memory accesses while processing the batch of events for any object.

At the implementation level, the queue is an array, and the elements of the array are (re)-used according to a circular buffer policy. Hence, when the events stored by the element corresponding with the epoch $E_i$ have all been processed (and $E_i$ ends) it can be reused to store new events, to be processed in the future, with timestamps falling in the epoch $E_{i+N}$.

However, it is possible that while some simulation object processes its events in the epoch $E_i$ along the execution of some thread, some new event is produced having timestamp beyond the upper limit of epoch $E_{i+N}$. In PARSIR each time one of these events is produced by a thread $THR_j$, we insert it into a per-thread fallback-list, relying on `__thread` head/tail pointers supported via Thread-Local-Storage (TLS). Each time an epoch ends, leading the time limit associated with the calendar to the end of the $E_{i+1+N}$ epoch, any thread $THR_j$ manages the passage of the events kept in its fallback-list to the calendar. Hence, the events still standing on the per-thread fallback-list are those not yet maintainable in the calendar, since their timestamp is still beyond the new limit related to the passage of epoch. Essentially, this type of organization enables parallelism in the management of the fallback-lists. At the same time, insertions and extractions from any individual fallback-list does not require synchronization, since the list is based on per-thread head/tail data that guarantee isolated traversals by any individual thread. In other words, PARSIR manages these fallback-lists in a manner somehow similar to kernel level per-CPU lists, like tasklet-lists in Linux.

Instead, any insertion into the calendar requires cross-thread synchronization. In fact, it requires the update of a linked list of event-buffers associated with some bucket. To make this synchronization step scalable, each calendar associated with any individual simulation object keeps a spinlock per bucket. Hence, spinlocks are organized in the following array

```
lock_buffer locks[OBJECTS][EPOCHS]
```

where `EPOCHS` corresponds to the number $N$ of epochs managed in the calendar, and the `lock_buffer` data type also embeds memory padding for separating the locks into different cache lines. This enables reducing the impact of actual RMW instructions (i.e. the Compare-and-Swap instruction) required for taking/releasing the locks on the cache coherency protocol used at the hardware level (e.g., MESI or MOESI [7]).

By the aforementioned structure of the event queue, when threads run the epoch $E_i$, no conflict will ever happen for accessing the $i$-th slot of the calendar associated with any individual object $o$. This is because, as we explained, object $o$ is exclusively managed by an individual thread $THR_j$, which took the object identifier in the workload distribution algorithm we have described in Section II-A. Hence, no other thread will try to extract events from that entry of the calendar associated with $o$, and at the same time, no other thread will attempt to insert any event in that same entry, since the lookahead $L$ leads newly produced events to be destined to future epochs, not $E_i$. This enables extractions from the calendar in the current epoch to be implemented with no need for acquiring the per-bucket spinlock. This is relevant since it allows the avoidance of the execution of RMW instructions (to get and release the lock), which have anyhow a cost related to cache lines usage and the Store-Buffer flush of the CPU the thread is running on.

Also, a conflict in the access to some bucket of the calendar associated with an object $o$ actually occurs only if two or more new events destined to object $o$ are contextually produced and injected in the queue via calls to `ScheduleNewEvent(...)` along multiple threads. However, for this to occur, we need that multiple other objects can have the object $o$ as the target of new events at the same wall-clock-time, and these new events need to have timestamps that require their insertion in the same bucket (hence they fall in the same epoch). These are factors that depend on the specific object-connectivity or timestamp generation distributions in the simulation model. But we can anyhow consider that our organization of the event queue, with such fine grain capability of control of synchronization on any individual per-object bucket, can provide an important support for conflict avoidance.

Overall, the combination of the above two points with the per-thread layout of the fallback-list can support with extremely high likelihood isolated (non-conflicting) operations on the engine level data structures used to implement the event-maintenance mechanisms. This leads PARSIR to be oriented to disjoint-access-parallelism, a fundamental property for scalable shared-memory oriented software platforms.

### C. NUMA Node Placement of Simulation Objects

The load distribution in PARSIR exploits the fact that each simulation object is hosted on a specific NUMA node of the underlying architecture. As said, this enables prioritizing local accesses in the NUMA architecture from the CPU where any thread is running.

At the same time, any simulation object is a generic data structure, which can be based on standard memory allocation services of the C programming language, like for example the ones offered by the `malloc` library, and on pointer-based linkage of the allocated areas. The actual choice is fully left to the application developer. At the same time, when the `ProcessEvent(...)` callback is activated by the PARSIR simulation engine, the first parameter is an integer `obj` communicating the identifier of the object for which an event is occurring. It can be used by the programmer to simply access an entry, e.g., in an array of `void*` pointers, for retrieving the address of the first memory chunk exploiting which the object state can be fully accesses, still via pointer-based linkage.

Now, the core point in PARSIR is to link memory allocations occurring when the `ProcessEvent(...)` callback is running for a given `obj` to the real memory allocation from the correct NUMA node where this object needs to be hosted.

To address this problem, we have exploited the following two solutions:

1) At simulation startup, PARSIR sets up an allocator for each simulation object, which is actually used when any memory allocation request is invoked while processing an event at that object.
2) The actual memory for serving the allocation requests is pre-allocated at the operating system level via the `mmap(...)` system call. At the same time, for each pre-allocated memory area, PARSIR exploits the `mbind(...)` system call for telling the operating system kernel that any mapped page in that zone needs to be materialized in RAM in a specific NUMA node.

We note that pre-allocation from the operating system is the de-facto standard approach used in common memory allocation libraries, like the standard `malloc` library implementation offered by GNU [8]. Hence, PARSIR does not contrast any best practice for memory allocation services.

The interception of memory allocation calls to the standard library is fully transparent since it is based on the `--wrap` linker facility. Hence, when a simulation object is processing an event, a call to, e.g., the `malloc(...)` API is simply routed to a call the allocator offered by PARSIR. In more detail, PARSIR offers a link-time flag that enables wrapping any standard library service that ultimately relies on dynamic memory allocation (e.g., the `strdup(...)` service).

At the same time, in order to know for which object we are actually allocating memory (hence what mapped area—what allocator instance—can be used for this allocation request, which as we explained is bound to a specific NUMA node), each thread in PARSIR keeps on a `__thread` variable called `current` the identity of the simulation object for which the `ProcessEvent(...)` callback has been started.

The overall organization of the allocator offered by PARSIR is definitely aimed at performance. In particular, the allocator for a given object is simply an array, with an element for each allocatable memory size, which is a power of 2 (by default the minimal allocatable size of memory, which we denote as MIN, is 32 bytes). Also, for each memory zone where allocations of a given size can occur for a specific simulation object, we keep an instance of the following data-structure type

```
typedef struct _area{
        void ** addresses;
        int top_elem;
        int count;
} area;
```

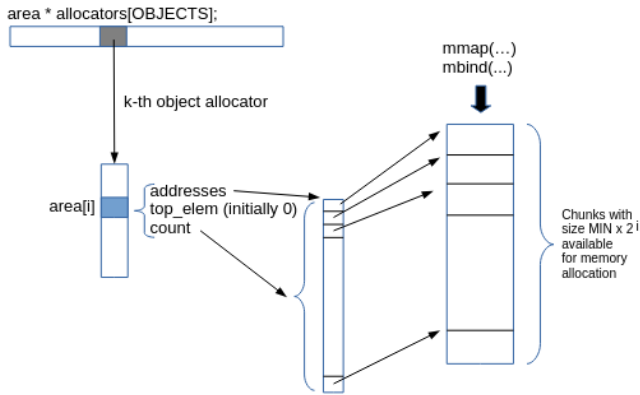

Fig. 1. Structure of the per-object NUMA oriented memory allocator.

TABLE I
DISCUSSED SIMULATION PLATFORMS

| Name | Technology | Type | Parallelism |
| --- | --- | --- | --- |
| SIMIAN [9] | Phyton | Gen. purpose | Yes - Conservative |
| Cunetsim [10] | CUDA | Cont. specific | Yes - Cons. (SIMD) |
| PARADISE++ [11] | C++ | Cont. specific | Yes - Optimistic |
| sPyNNaker [12] | C | Cont. specific | Yes - Conservative |
| DESP-C++ [13] | C++ | Gen. purpose | No |
| ROOT-Sim [14] | C | Gen. purpose | Yes - Optimistic |
| ROSS [15] | C | Gen. purpose | Yes - Optimistic |
| JavaSim [16] | Java | Gen. purpose | No |
| ARTIS [17] | C | Gen. purpose | Yes - Conservative |
| USE [18] | C | Gen. purpose | Yes - Optimistic |
| PARSIR | C | Gen. purpose | Yes - Conservative |

The fields in this data structure are used for the following purposes: addresses points to an array of void* elements that record the addresses of chunks deliverable upon allocation requests; top_elem is the index of the array keeping track of the element to be used for the next memory allocation request; count is the size of the array of void* elements. In Figure 1, we show the structuring of the NUMA-oriented allocator, related to the above description.

This data structure supports the implementation of a stack allocator where an allocation is simply carried out by executing the following statement (based on a preceding setup of the pointer addresses to the corresponding field of the target area for the allocation) "return addresses[top_elem++]" and a release of a chunk at address addr (like when free(addr) is called by the application code and is transparently intercepted by the PARSIR engine) is simply carried out via the following update "addresses[--top_elem] = addr". At the same time, if the memory currently managed by the allocator gets exhausted, then a simple realloc of the addresses array can be executed after having pre-reserved other memory for the same simulation object (and after again setting the correct NUMA placement via mbind(...)). The reallocated array of addresses can be then updated with the new chunk addresses belonging to the newly mapped area.

A hashing mechanism—exploiting that mapping relies on the MAP_FIXED flag—is used to determine what mapped area a chunk being released belongs to, in order to immediately access the correct entry of the array of area structures.

A final important note related to resource usage is that the logical pages belonging to mapped zones are not really hosted in any RAM frame unless their content (hence some memory chunk they host) is actually read/write accessed by the simulation object after having requested the memory allocation. This is the classical materialization of the logical page in RAM carried out by the operating system kernel through page-fault upon the initial access to the mapped page. Such an advantage is achieved also thanks to the avoidance of metadata for the memory manager into the head portion of each usable chunk—in fact this avoids the need for initializing these metadata, which would otherwise lead to logical page materialization in RAM because of the write operations on the metadata. This feature can help saving RAM even in scenarios with (very) large mapped zones of memory.

III. RELATION WITH THE LITERATURE

In this section we provide a comparative discussion focusing our attention on a set of (P)DES simulators/packages that are taken from the literature. The list we discuss is not an exhaustive list of what can be found in the literature, but rather it is biased to packages that appear to already include some optimizations, or have some peculiarity, compared to other literature solutions. For this comparative discussion we rely on Table I, where we list the simulation packages we include in the discussion, synthesizing a few relevant aspects useful for the analysis (e.g., if the simulation package is general purpose or is context specific).

SIMIAN [9] is based on Python, and supports the parallel execution of models via the exploitation of message passing at the level of the MPI layer. One of the main objectives of SIMIAN has been the one of porting the JIT (Just-in-Time) compilation technique to the parallel simulation community. Our engine PARSIR is substantially different since it relies on the C language, on optimized architectural design and implementation for shared-memory machines (with no reliance on any message-passing layer), and directly tackles aspects inherent to the hardware-level setup.

Cunetsim [10] is a simulation framework suited for exploiting CUDA and GPU-computing. It has been proposed for simulating packet-level networks. Even though it exploits the CPU for a few tasks, it is not suited for the exploitation of large scale parallel machines with multiple CPUs and multiple

NUMA nodes, as instead PARSIR does. Also, PARSIR is essentially a general purpose simulation environment with no binding to specific classes of simulation models.

PARADISE++ [11] has been explicitly designed for simulating post-Moore HPC computing systems, in order to facilitate their design and evolution. It exploits the Devastator simulation engine [19], which supports optimistic parallel execution. Also, it is essentially based on C++ technology, and offers system-software oriented optimizations, like non-blocking data structures. PARSIR also offers similar optimizations, since it exploits the lock-free approach based on RMW instructions to distribute the workload across threads, and high likelihood of disjoint-access-parallelism, while also avoiding abort/retry that is typical of non-blocking data management algorithms (see, e.g., [20]). Furthermore, PARSIR embeds NUMA optimizations. Beyond this, a core difference is that PARADISE++ is based on optimistic synchronization. Hence, it requires the employment of mechanisms for managing rollbacks in case of the occurrence of causality violations. PARSIR does not require this type of support, therefore it offers a simplified management of simulation code, neither requiring the construction of reverse event handlers [21] or the support for checkpointing [22]–[24] for making it reversible. Furthermore, PARSIR is general purpose, and is not limited to a specific class of simulation models.

sPyNNaker [12] is a project that offers a machine relying on ARM technology and the support for C-based implementation of simulation models of PyNN-defined spiking neural networks. It is based on timer events used to raise the execution of callbacks that perform neuron state updates and manage pipelined event-driven spike processing. Differently from PARSIR, this package has applications in a unique class of simulations (the one of neural-network models). Also, it is designed for a specific hardware setup, while PARSIR is essentially portable across any Posix operating system, and can be hosted on both bare-metal and virtual machines.

As for C++ simulation code, DESP-C++ [13] offers a very simple support for its development. However, this package does not support parallelism in the execution of a model, and is essentially an object-oriented engine for the management of a calendar of events, and their correct CPU scheduling along the virtual time axis. Overall, even though the core technology used has some relation with the one adopted in PARSIR (i.e., C++ vs C), the two packages are substantially different.

The ROOT-Sim package [14] offers the possibility to run C-based simulation models on top of parallel/distributed environments, according to the optimistic synchronization paradigm. Its development releases have provided various optimizations for handling scalability and high performance of the execution of simulation models, like for example NUMA optimization. However, this package is essentially different from PARSIR, since it needs to include the support for reversibility of the simulation code execution. Additionally, specific optimizations it has embedded along its timeline are essentially orthogonal to the ones employed by PARSIR. Just to mention a few, the NUMA optimization of ROOT-Sim presented in [25] requires migration of virtual pages across the NUMA nodes, while PARSIR does not require page migrations, which have a cost since they are executed in a synchronous manner upon their invocation at the operating system level. Additionally, the batch execution of the events of an individual object—adopted by PARSIR for improving caching efficiency—is not exploited in ROOT-Sim, which instead always schedules along a thread the minimal timestamp event of some object currently bound to that same thread. Overall, ROOT-Sim switches across the different objects in a time interleaved manner—with possible impact on locality reduction—depending on the event timestamps.

Similar considerations can be made for the ROSS package [15], since it is still based on optimistic parallel/distributed execution of discrete event simulation models developed with the C programming language—while PARSIR relies on the conservative synchronization scheme[1]. In ROSS, the interactions across simulation objects are supported via MPI, hence relying on a pure message-passing paradigm, while PARSIR is explicitly devised for shared-memory machines, and adopts a data-structure sharing, supported via non-blocking and high-likelihood disjoint access parallel algorithms, with automatic maximization of locality in the accesses to object states on NUMA machines. At the same time, no support for NUMA migration is offered by ROSS, which still leaves the accesses to the simulation object served according to the distance between CPUs and NUMA nodes established when materializing the simulation objects' states (e.g., the first-touch policy at the operating system level). Also, similarly to ROOT-Sim, the ROSS package does not adopt batch processing of events at an individual simulation object, while PARSIR does.

JavaSim [16] is a package devised for the development of discrete event simulation models using object oriented programming. It is classically structured in order to make the simulation engine exploit specific capabilities or configurations of the underlying computing system (e.g., operating system) through common libraries supporting the target object oriented language (Java or C++). Contrarily, PARSIR is by design constructed in order to directly exploit this kind of system-level configuration (e.g., operating system services) or hardware specific features (e.g., the atomic RMW support of memory locations offered by the ISA). Also, in JavaSim parallelism (in the sense of PDES) is not offered. In fact, in JavaSim multiple processes are executed in pseudo-parallel mode, i.e., only one process executes at any instant of real time, even though many processes may execute concurrently at any instant of simulation time—having events with that timestamp. Contrarily, PARSIR is fully oriented to parallel execution of discrete event simulation models.

ARTIS [17] is a framework for conservative execution of simulation models relying on the C programming language on parallel/distributed computing systems. It exploits R-UDP/IP or TCP/IP for distributed communication. It supports migrating objects across threads, being it still linked to an execution model where a (temporary) binding between threads and object is established. PARSIR exploits a different approach where in each epoch (time-step) of the execution of the simulation, each thread can steal workload to any other, favoring at the same time local NUMA accesses.

---

[1]ROSS also supports conservative synchronization, for example exploitable for simplicity in the application development phase. However, as also pointed out in [26] its performance/scalability targets are reached essentially through the offered optimistic synchronization support.

USE [18] has been designed for multi-processor shared-memory machines. It avoids the usage of any message-passing layer and fully relies on shared data structures at the simulation engine level. In particular, it relies on a non-blocking algorithm [27] for managing an event pool fully shared across the worker threads. It makes the thread always process the available events with timestamps as close as possible to the commit horizon of the optimistic simulation run, with possible advantages in terms of reduction of the incidence of causality violations. With this solution the threads continuously switch their execution across different objects, which does not favor locality. Hence, some optimizations have been provided in order to limit the impact of this switch on locality degradation [28]. These are anyhow based on the usage of object migrations across the NUMA nodes in the underlying architecture, which is instead not used in PARSIR, since it prioritizes local NUMA node accesses at the work-stealing level.

Overall, PARSIR looks to provide a few performance-oriented architectural/technical solutions that are different, compared to what already provided by literature packages, which make it an attractive alternative.

## IV. BENCHMARK EXPERIMENTS

### A. Test-bed

We report the results of some experiments in order to provide the reader with indications on the execution profile of PARSIR. In the experiments, we decided to rely on the usage of the common PHOLD benchmark [29]. This benchmark has been used in various configurations by many studies in the area of PDES (see, e.g., [28], [30]–[33]) for assessing the effectiveness of innovative solutions and innovative simulation architectures/platforms. It is therefore a well known and widely exploited benchmark in the PDES community. Also, PHOLD is typically implemented as test-bed in various parallel discrete event simulation packages. Hence, its usage also enabled us to perform a comparative analysis with a few of the packages available in the literature.

Although it can generate workloads with extremely diversified execution profiles, its structure is relatively simple. It has the following set of baseline parameters that can be configured:
- the number $O$ of simulation objects;
- the number $M$ of initial events scheduled at any object;
- the lookahead $L$.

When an event is processed by an object, a new event with timestamp in the future is generated. The timestamp value depends on a given distribution of timestamp increment (e.g., exponential) and the lookahead. This new event can be routed to any object in the simulation model, according to a uniform or non-uniform distribution.

The cross-scheduling of simulation events among any couple of objects in PHOLD gives rise to the general scenario where any part of the simulation model can be related to/affected by any other part—in terms of occurrence of things along the simulation time axis. This is a challenge for PDES synchronization algorithms and for the associated platforms [1]. In fact, PHOLD has initially had a relevant role in the assessment of new algorithm proposals in both conservative and optimistic synchronization.

However, this benchmark has been extended (see, e.g., [34]) in order to enable the assessment of how other aspects inherent to the PDES platform—like for example the support for locality of the operations occurring at the level of the simulation software—can affect performance. For this reason, the state of an individual object in PHOLD has been devised as a generic data structure (e.g., memory contiguous or pointer linked) of size $S$ on which the event performs a set of read/write operations and reallocates a percentage $P$ of the whole state size.

In our experiments, we used an implementation of the benchmark based on multiple linked lists that represent the state of the objects. In particular, we used two different lists of chunks that are requested to be 32 or 64 bytes upon their allocation. Each node on the lists keeps the pointer to the next node—made of 8 bytes—plus the payload area, where the remaining bytes of the chunk associated with the node are simply used for memory copy operations miming the ones implemented in various real-world models. As an example, for the simulation of personal communication systems, some study uses a list of busy channels [28] which is scanned when events occur in order to read/write the corresponding information (e.g., the current power usage for the ongoing communication on the channel, which is used to setup the one related to additional incoming calls for reaching the target Signal-to-Interference Ratio [35]). Also, dynamic release/allocation of list elements takes place while processing the simulation events, in order to mimic the scenario where an event can have impact on the layout of the simulation object state.

Table II summarizes the range of values we have employed for the PHOLD parameters. We note that moving the value of $O$ (number of simulation objects) in the interval [1024, 8192], in combination with different $S$ (object state-size) configurations provides the support for determining the outcomes of the execution of the simulation package for an ample memory demand interval. Actual access operations to the lists of chunks in the state layout of an object have been configured to touch 1/32 of the total number of elements in the lists. Furthermore, moving the value of $M$ in the interval [10, 1000] allows the mimic of simulation dynamics with definitely different densities of events along the simulation time axis. Also, the variation of the value of $P$ in the interval [0.1%, 0.4%] enables generating models where a simulation object relocates (via `malloc`/`free` calls) different tens of list-nodes in its state. Finally, moving $L$ in the interval [1/10, 1.0] enables very disparate impacts in terms of synchronization, since the volume of events that can be actually executed while being causally consistent before passing to new epochs along the simulation time axis changes drastically.

TABLE II
ADOPTED VARIATION INTERVALS OF THE PHOLD PARAMETERS

| PHOLD parameter | Variation interval |
| --- | --- |
| $O$ (number of objects) | [1024, 8192] |
| $M$ (number of initial events per object) | [10, 1000] |
| $S$ (state size per object - list nodes) | [4000, 16000] |
| $P$ (per event reallocated object-state fraction) | [0.1%, 0.4%] |
| $L$ (lookahead of the model - sim. time units) | [1/10, 1.0] |

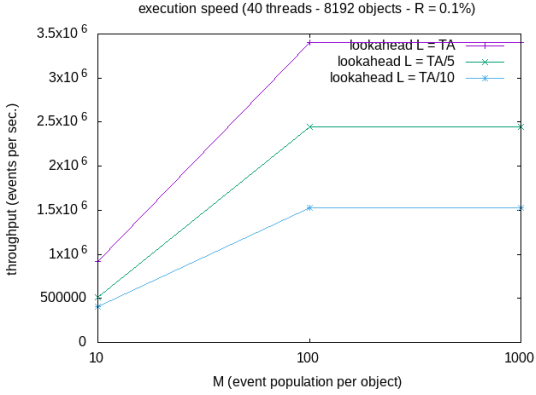

Fig. 2. Variation of the execution speed vs $L$ and $M$

### B. Computing Platform

Our experiments have been run on top of a machine equipped with 2 Intel(R) Xeon(R) Silver 4210R processors working at 2.40GHz. Each processor has 10 CPU-cores and 20 Hyper-Threads (which we simply refer to as CPUs). Hence the total number of CPUs available using the two processors is 40. The machine is equipped with 32GB of RAM organized in 2 different NUMA nodes, each one hosting 16 GB of memory. Each CPU sees a 32KB upper level cache and a 14080KB lower level cache. The machine runs the Linux operating system, in particular the Ubuntu 18.04 release. Version 7.5.0 of gcc has been used to compile the packages that have been included in this experimental assessment.

### C. Execution Outcomes

A first set of experiments has been done in order to determine how stable is the performance of PARSIR when fixing the simulation model size, in terms of number of objects $O$ and their state size $S$, and changing parameters inherent to the simulation model execution, which can impact the epoch based approach adopted by PARSIR. For this experiment, we set $O$ to the value 8192, $S$ (the number of nodes in the lists kept in the state of any simulation object) to the value 16000 and $P$ (the percentage of the simulation object state reallocated when executing an event) to the value 0.1%. Hence we relied on the maximal model size listed in Table II. At the same time, we varied $M$ between 10 and 1000, and the lookahead $L$ between 0.1 and 1.0 of the average timestamp increment, which we denote as $TA$ in the plots—the distribution of the timestamp increment has been set to exponential. Also, the runs have been executed with 40 threads. Each run was set to last 1 minute, and we report the average value of the event throughput (number of simulation events processed per wall-clock-time unit) computed by relying on 10 samples (the differences among the samples were under 3%).

The results for this setup are reported in Figure 2. By the outcomes we see how PARSIR provides extremely stable performance when the event population per simulation object $M$ is larger than or equal to 100. This stability is noted independently of the lookahead value. This is a relevant outcome indicating that models with significantly different lookahead values can be processed effectively by PARSIR, especially when the density of events along the virtual time axis is non-minimal. In any case, the stable value reached by any configuration can represent a kind of "optimal" performance. achievable for that specific model settings (in terms of model size and lookahead value).

A second set of experiments has been devoted to assess a core aspect of any parallel execution framework, namely strong scalability. In particular, we tested the effectiveness of PARSIR fixing the model size to a given value and then scaling up the number of threads used to run the model. We set $O$ to 8192, $S$ to 16000 and $P$ to 0.1%. At the same time we varied the lookahead $L$ between 0.1 and 1.0 of $TA$ and the value $M$ of the event population per object from 10 to 1000. Also, the scalability study has been performed changing the number of used threads from 10 to 40, with step 10 increment at each variation. We report the throughput of executed simulation events computed as the mean value over 10 different samples (minimal difference has been observed between the different samples).

The results are shown in Figure 3. We can see how PARSIR enables acceleration when increasing the number of used threads, with a linear growth of the execution speed when using larger thread counts. This happens independently of the event population setup—namely the setup of the parameter $M$. At the same time, the reduced frequency of occurrence of the epoch barrier along the wall-clock-time when $M$ is larger makes the scalability curve to stay higher. Also, an interesting observation comes out for the case of lookahead $L$ set equal to $TA$. In this case, the scalability curve with $M$ set to 1000 is lower than the one with $M$ set to 100. The reason for this behavior is that when using $L$ equal to $TA$ there is a higher density of events that fall in the same epoch and that can target the same simulation object. This is particularly true when running with higher event population. Hence, there is a decrease of the likelihood of disjoint access parallelism in the different buckets of the event queues used by PARSIR. However, for large values of the model lookahead, this aspect can be easily faced by simply running PARSIR with an epoch size set to a fraction of the lookahead.

Finally, we report data related to the behavior of PARSIR when changing the simulation model size, in particular when scaling the model size from 1024 to 8192 simulation objects. For this scaling experiment, we adopted a different value of $P$, setting it to 0.4%. Also, similarly to the above experiments, we still varied the lookahead $L$ from 0.1 to 1.0 of $TA$ and $M$ from 10 to 1000. Also, we scaled the model size while keeping the thread count fixed to the maximum value 40, and we report the average event throughput over 10 samples.

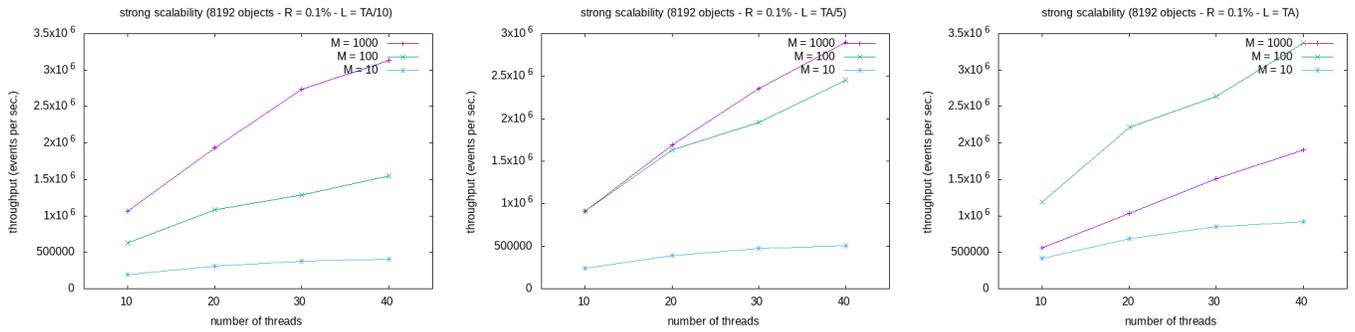

Fig. 3. Variation of the execution speed vs the number of threads (strong scaling)

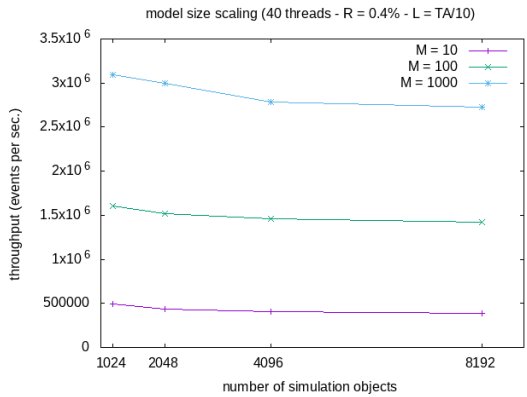

Fig. 4. Variation of the execution speed vs the model size

The experimental data are shown in Figure 4. As one likely expects from a well structured/realized parallel processing platform, once fixed the number of used threads, the actual performance when increasing the model size is essentially flat. This indicates how the data structures used by PARSIR for handling increasing numbers of simulation objects (from 1024 to 8192) have a relatively low impact in terms of reduction of the execution locality. At the same time, the lower throughput curve observed when decreasing $M$ is just due to the reduction of the density of events per execution epoch, which in turn necessarily leads to processing smaller sets of events per-object in a batch. This leads to reduced locality and reduced cache-level effectiveness, which add their impact to the one of the more frequent barrier occurrence.

### D. Comparison with other Packages

We also performed a comparison with a few packages available in the literature. The selection of the simulation packages tested against PARSIR is based on the outcomes synthesized in Table I and on the support really offered by the packages. In particular, we selected all the packages that:

- are based on C technology,
- are devised to be general purpose,
- support parallelism and already offer an implementation of the selected benchmark PHOLD,
- support model programmability relying on the standard library, in particular on the `malloc` and `free` services for dynamically changing the layout of the state of the simulation objects; we feel this aspect is important for an ample possibility of usage.

Based on the above criteria, the final list of packages for which we provide experimental data includes our package PARSIR plus ROOT-Sim and USE—each of the other packages does not comply with one or more of the above listed requirements. The shortlisted packages all offer the same API to the simulation application code. Hence, in this comparison we simply exploited a uniquely developed configuration of the PHOLD benchmark parameterized for this experimental study and simply used it according to a copy/paste approach of code blocks applied to the PHOLD skeletons offered by the packages. This also contributed to fairness in the comparison. Additionally, one interesting point in this shortlist of compared packages is that it provides solutions that mix conservative and optimistic synchronization approaches. As a final preliminary note, for ROOT-Sim and USE we selected the package configuration that is at current date offered by the package mainline. Material provided by other branches is therefore not used in this experimental assessment.

For this comparison we performed a strong scaling experiment with $O$ set to 2048 and $S$ set to 4000, and $P$ set to 0.4%. This is an intermediate model size, in the set of the configurations listed in Table II, which has been selected in order to avoid excessive memory usage that, as well known in the literature [21], [36], could be more adverse for the optimistic synchronization approaches used in ROOT-Sim and USE. Additionally, again for not disfavoring ROOT-Sim and USE vs PARSIR, we used the minimal value of the lookahead $L$, which has been set to 1/10 of the average timestamp increment $TA$. Also, we used the minimal event population value expressed in Table II which, as discussed before, can be more adverse for PARSIR. In particular, we set the value of $M$ to 10. For this experiment we still report the average event throughput while scaling the number of used threads.

The results are reported in Figure 5. They show how

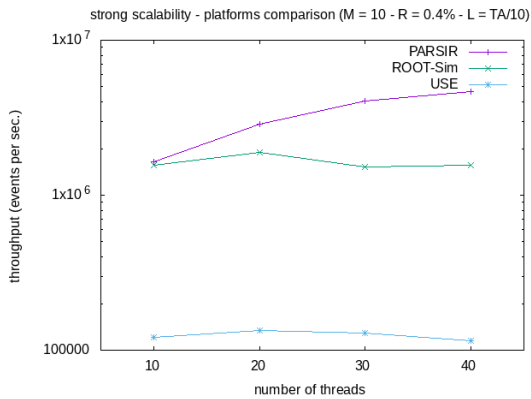

Fig. 5. Comparison with other packages - Strong scalability

PARSIR offers a better scalability curve in terms of both its height and its shape. In particular, both ROOT-Sim and USE do not allow scaling up the performance when running with more than 20 threads (compared to scenarios when they run with less than or with 20 threads). Also, while the maximal execution speed provided by PARSIR is of the order of $3.5 \times 10^6$ events per second, the other two packages show a significantly lower peak value of their event throughput. In fact, USE shows the order of $1.4 \times 10^5$ events per second as its peak throughput, while ROOT-Sim shows peak throughput of $1.8 \times 10^6$ events per second.

## V. CONCLUSIONS

In this article, we have presented PARSIR (PARallel SImulation Runner), an open source package for developing and running discrete event simulation models on top of shared-memory multi-processor machines. PARSIR embeds a set of technical solutions that allow it to achieve high performance when running simulation models on the target computing platforms. In particular, it embeds solutions oriented to cache effectiveness, NUMA deploy effectiveness, non-blocking RMW based assignment of the workload to threads and (high likelihood of) disjoint-access-parallelism. The achievement of high performance and scalability has been shown in this article by relying on the well known PHOLD benchmark, and by running different configurations of this benchmark on top of a shared-memory machine equipped with two NUMA nodes and 40 CPUs (Hyper-Threads). We have also reported data for a comparison of PARSIR against other—research oriented—parallel discrete event simulation packages available for free download. Beyond its immediate usability, we think that the design/implementation of PARSIR includes a set of solutions that can result of interest for readers working in the area of parallel computing. Also, the availability of PARSIR as an open source package puts all these solutions immediately available for any experimentation or additional idea for research in the area of parallel discrete event simulation platforms.

## ACKNOWLEDGMENT

This paper has been partially supported by the Italian MUR PRIN 2022 Project: Domain (Grant #2022TSYYKJ) financed by NextGenEu.

## REFERENCES

[1] R. Fujimoto, "Parallel discrete event simulation," *Commun. ACM*, vol. 33, no. 10, pp. 30–53, 1990.

[2] E. Deelman, R. Bargodia, R. Sakellariou, and V. Adve, "Improving lookahead in parallel discrete event simulations of large-scale applications using compiler analysis," in *Proceedings of the 15th Workshop on Parallel and Distributed Simulation*, 2001, pp. 5–13.

[3] M.-K. Chung and C.-M. Kyung, "Improving lookahead in parallel multiprocessor simulation using dynamic execution path prediction," in *Prooceedings of the 20th Workshop on Principles of Advanced and Distributed Simulation*, 2006, pp. 11–18.

[4] J. Wang, Z. Dong, S. Yalamanchili, and G. F. Riley, "FNM: an enhanced null-message algorithm for parallel simulation of multicore systems," *ACM Trans. Model. Comput. Simul.*, vol. 26, no. 2, pp. 11:1–11:26, 2016.

[5] R. M. Fujimoto, "Research challenges in parallel and distributed simulation," *ACM Trans. Model. Comput. Simul.*, vol. 26, no. 4, pp. 22:1–22:29, 2016.

[6] "PARSIR (PARallel SImulation Runner)." [Online]. Available: https://github.com/FrancescoQuaglia/PARSIR.git

[7] J. L. Hennessy and D. A. Patterson, *Computer Architecture, Sixth Edition: A Quantitative Approach*, 6th ed. San Francisco, CA, USA: Morgan Kaufmann Publishers Inc., 2017.

[8] "The gnu allocator." [Online]. Available: https://www.gnu.org

[9] "Simian parallel discrete event simulator." [Online]. Available: https://pujyam.github.io/simian/

[10] B. B. Romdhanne and N. Nikaein, "Cunetsim: a new simulation framework for large scale mobile networks," in *Proceedings of the International ICST Conference on Simulation Tools and Techniques*, 2012, pp. 217–219.

[11] "Paradise++ large scale optimistic synchronization based simulation of post moore systems." [Online]. Available: https://crd.lbl.gov/divisions/amcr/computer-science-amcr/cag/research/paradise/

[12] "spynnaker: A software package for running pynn simulations on spinnaker." [Online]. Available: https://www.frontiersin.org/articles/10.3389/fnins.2018.00816/full

[13] J. Darmont, "DESP-C++: A discrete-event simulation package for C++," *CoRR*, vol. abs/1611.09170, 2016. [Online]. Available: http://arxiv.org/abs/1611.09170

[14] A. Pellegrini and F. Quaglia, "The ROme OpTimistic Simulator: A tutorial," in *Euro-Par 2013: Parallel Processing Workshops*, ser. Lecture Notes in Computer Science, vol. 8374. Springer, 2013, pp. 501–512.

[15] "Ross." [Online]. Available: https://ross-org.github.io/

[16] "Javasim." [Online]. Available: https://github.com/nmcl/JavaSim

[17] L. Bononi, M. Bracuto, G. D'Angelo, and L. Donatiello, "Scalable and efficient parallel and distributed simulation of complex, dynamic and mobile systems," in *Proceedings of the 2005 Workshop on Techniques, Methodologies and Tools for Performance Evaluation of Complex Systems*, 2005.

[18] M. Ianni, R. Marotta, D. Cingolani, A. Pellegrini, and F. Quaglia, "The ultimate share-everything PDES system," in *Proceedings of the 2018 ACM SIGSIM Conference on Principles of Advanced Discrete Simulation*, 2018, p. 73–84.

[19] "Devastator parallel discrete event simulation runtime (devastator) v1.0." [Online]. Available: https://www.osti.gov/biblio/1820931

[20] R. Marotta, M. Ianni, A. Pellegrini, and F. Quaglia, "NBBS: A non-blocking buddy system for multi-core machines," *IEEE Trans. Computers*, vol. 71, no. 3, pp. 599–612, 2022.

[21] C. D. Carothers, K. S. Perumalla, and R. Fujimoto, "The effect of state-saving in optimistic simulation on a cache-coherent non-uniform memory access architecture," in *Proceedings of the 31st Winter Simulation Conference*, 1999, pp. 1624–1633.

[22] A. Pellegrini, R. Vitali, and F. Quaglia, "Autonomic state management for optimistic simulation platforms," *IEEE Trans. Parallel and Distributed Syst.*, vol. 26, no. 6, pp. 1560–1569, 2015.

[23] D. West and K. S. Panesar, "Automatic incremental state saving," in *Proceedings of the 10th Workshop on Parallel and Distributed Simulation*, W. M. Loucks and B. R. Preiss, Eds., 1996, pp. 78–85.

[24] F. Quaglia and A. Santoro, "Nonblocking checkpointing for optimistic parallel simulation: Description and an implementation," *IEEE Trans. Parallel and Distributed Syst.*, vol. 14, no. 6, pp. 593–610, 2003.

[25] A. Pellegrini and F. Quaglia, "NUMA Time Warp," in *Proceedings of the ACM SIGSIM Conference on Principles of Advanced Discrete Simulation*, 2015, p. 59–70.

[26] M. Mubarak, C. D. Carothers, R. B. Ross, and P. H. Carns, "Enabling parallel simulation of large-scale HPC network systems," *IEEE Trans. Parallel and Distributed Syst.*, vol. 28, no. 1, pp. 87–100, 2017.

[27] R. Marotta, M. Ianni, A. Pellegrini, and F. Quaglia, "A conflict-resilient lock-free linearizable calendar queue," *ACM Trans. Parallel Comput.*, vol. 11, no. 1, pp. 4:1–4:32, 2024.

[28] F. Montesano, R. Marotta, and F. Quaglia, "Spatial/temporal locality-based load-sharing in speculative discrete event simulation on multi-core machines," in *Proceedings of the SIGSIM Conference on Principles of Advanced Discrete Simulation*, 2022, pp. 81–92.

[29] R. M. Fujimoto, "Performance of Time Warp Under Synthetic Workloads," in *Proceedings of the Multiconference on Distributed Simulation*, 1990, pp. 23–28.

[30] P. D. Barnes, C. D. Carothers, D. R. Jefferson, and J. M. LaPre, "Warp speed: executing time warp on 1, 966, 080 cores," in *Proceedings of the ACM SIGSIM Conference on Principles of Advanced Discrete Simulation*, 2013, pp. 327–336.

[31] C. D. Carothers and R. Fujimoto, "Efficient execution of time warp programs on heterogeneous, NOW platforms," *IEEE Trans. Parallel and Distributed Syst.*, vol. 11, no. 3, pp. 299–317, 2000.

[32] P. Andelfinger, T. Köster, and A. M. Uhrmacher, "Zero lookahead? zero problem. the window racer algorithm," in *Proceedings of the ACM SIGSIM Conference on Principles of Advanced Discrete Simulation*, 2023, pp. 1–11.

[33] S. Rahman, N. B. Abu-Ghazaleh, and W. A. Najjar, "PDES-A: accelerators for parallel discrete event simulation implemented on fpgas," *ACM Trans. Model. Comput. Simul.*, vol. 29, no. 2, pp. 12:1–12:25, 2019.

[34] R. Vitali, A. Pellegrini, and F. Quaglia, "Benchmarking memory management capabilities within ROOT-Sim," in *Proceedings of the 13th IEEE/ACM International Symposium on Distributed Simulation and Real Time Applications*, 2009, pp. 33–40.

[35] S. Kandukuri and S. Boyd, "Optimal Power Control in Interference-Limited Fading Wireless Channels with Outage-Probability Specifications," *IEEE Transactions on Wireless Communications*, vol. 1, no. 1, pp. 46–55, 2002.

[36] S. R. Das and R. Fujimoto, "Adaptive memory management and optimism control in time warp," *ACM Trans. Model. Comput. Simul.*, vol. 7, no. 2, pp. 239–271, 1997.